\documentclass{article}
\usepackage{graphicx} 
\usepackage[letterpaper,top=2cm,bottom=2cm,left=3cm,right=3cm,marginparwidth=1.75cm]{geometry}

\usepackage{upgreek, color,amsfonts}
\usepackage{amsmath, hyperref,amssymb, bbm,breqn,color, comment,authblk}

\usepackage[numbers,sort&compress]{natbib}
\makeatletter
\def\BibitemShut#1{\unskip\csname bibitem#1\endcsname}
\makeatother

\def\Tr{{\rm Tr}}
\def\Ric{Ricci}

\title{\Large Towards gauge independence in asymptotically safe quantum gravity}
\author[1]{Kevin Falls}
\author[2]{Renata Ferrero}
\author[1,3,4,5]{Giovanni Oglialoro}
\affil[1]{\small Instituto de Física, Facultad de Ingeniería, Universidad de la República, J.H.y Reissig 565, 11300 Montevideo, Uruguay}
\affil[2]{\small Institute for Quantum Gravity, FAU Erlangen – Nürnberg, Staudtstr. 7, 91058 Erlangen, Germany}
\affil[3]{\small INAF – Osservatorio Astrofisico di Catania, Via S. Sofia 78, 95123 Catania, Italy}
\affil[4]{INFN, Sezione di Catania, Via S. Sofia 64, 95123 Catania, Italy}
\affil[5]{Dipartimento di Fisica e Astronomia, Universit\`a di Catania, Via S. Sofia 64, 95123, Catania, Italy}
\date{}

\begin{document}

\maketitle

\begin{abstract}
We study gauge dependence in proper-time renormalisation group flows for asymptotically safe quantum gravity. Working in an essential scheme, we use field redefinitions to separate redundant off-shell contributions from the on-shell running of physical couplings. We consider two approximations: in the first, we work on general backgrounds and project onto all terms with up to four derivatives, neglecting boundary terms; in the second, we work on a maximally symmetric background and retain all orders in the Ricci scalar. Although the flow equation depends on the gauge parameters, this dependence can be cancelled order by order once the redundant terms are absorbed by field redefinitions. In the four-derivative approximation, the dependence on the retained gauge parameter drops out of the beta function for Newton's constant to third order in the coupling. On the sphere, the flow becomes independent of both parameters of the general gauge to all curvature orders, and all orders in the coupling. Crucially, this cancellation depends on the choice of regulator. We verify the mechanism explicitly at one loop, where the gauge-dependent fluctuation contributions are cancelled by the ghost sector on shell. The resulting essential flow displays a gauge-independent non-Gaussian fixed point, supporting the interpretation that universal information in quantum gravity is encoded in the on-shell essential sector.
\end{abstract}

\tableofcontents

\section{Introduction}

A central requirement for any quantum field theoretic treatment of gravity is that physical results should be independent of the gauge fixing used to define the theory. In the covariant path integral, gauge fixing is introduced in order to invert the quadratic fluctuation operator and define the Faddeev--Popov determinant, but the associated gauge parameters are auxiliary and cannot affect physical observables. In practice, most renormalisation group constructions for gravity are formulated in terms of effective actions away from the equations of motion, and this makes the separation between physical and unphysical information highly non-trivial. Off-shell effective actions generally depend on the gauge fixing, on the parametrisation of the metric fluctuation, and on the particular regularisation scheme used to implement the flow. As a consequence, a major conceptual issue in asymptotic safety~\cite{Weinberg:1976xy,Reuter:1996cp,Percacci:2017fkn,Bonanno:2020bil} is to understand to what extent the fixed-point structure extracted from approximate flow equations reflects universal physics rather than artifacts of the off-shell description.

This issue has been recognised for a long time in functional approaches to gravity~\cite{Reuter:1996cp,Wetterich:1992yh,Morris:1993qb,Ellwanger:1993mw,Pawlowski:2005xe}. Some attempts to formulate gauge invariant flows have been put forward \cite{Litim:1998nf,Morris:1998kz,Morris:1999px,Morris:2000fs,Arnone:2002cs,Morris:2016nda,Wetterich:2016ewc,Falls:2020tmj,Sonoda:2020vut,Ihssen:2025cff}. However, the effective average action and related proper-time flows~\cite{Bonanno:2004sy,Bonanno:2000yp,Mazza:2001bp,Bonanno:2019ukb} are naturally formulated for gauge-fixed actions on a background geometry, and the regulator itself typically acts differently on different fluctuation sectors. Even when diffeomorphism invariance is restored in the physical limit, intermediate stages of the flow need not display manifest gauge independence. In finite-dimensional truncations, this can lead to a residual dependence of beta functions and critical exponents on gauge parameters or on the details of the field parametrisation~\cite{Falkenberg:1996bq,Ohta:2016npm,Ohta:2016jvw,Gies:2015tca,Nink:2014yya,Dou:1997fg,Bonanno:2025tfj,DAngelo:2025yoy}. Whereas in perturbation theory gauge dependence is known to cancel order by order for on-shell observables \cite{Parker:2009uva}, here we move towards the non-perturbative regime, where this cancellation is no longer automatic. Such a dependence is not necessarily a sign of inconsistency, but it obscures the identification of the  universal sector of the theory and raises the question of how the renormalisation group flow should be projected onto physically meaningful couplings. 

A particularly useful viewpoint on this problem has recently emerged from the on-shell perturbative formulation of asymptotic safety developed in \cite{Falls:2024noj}. There, the renormalisation of Einstein gravity is studied in a subtraction scheme based on dimensional regularisation \cite{Kluth:2024lar,Beretta:2026zcy,Beretta:2026avu}, and the flow is organised directly around essential couplings~\cite{Baldazzi:2021ydj,Baldazzi:2021orb,Knorr:2023usb,Knorr:2022ilz}. Working in a one-loop improved approximation, the key point is that field redefinitions~\cite{Kamefuchi:1961sb,Chisholm:1961tha,Kallosh:1972ap,Efimov:1972juh,Tyutin:2000ht,Arzt:1993gz} can be used to remove off-shell contributions to the renormalisation group equations, thereby isolating the running associated with on-shell data. Within this framework, the one-loop beta function for Newton's coupling can be mapped to the corresponding result obtained with a proper-time cutoff, while the dependence on the parametrisation of the metric fluctuation is removed order by order within the approximations considered. Moreover, by extending the analysis to all orders in the scalar curvature, one obtains a beta function for Newton's coupling that is independent of the parametrisation and vanishes at the Reuter fixed point~\cite{Reuter:1996cp,Reuter:2001ag}. Much of the apparent scheme and parametrisation dependence of gravitational flows can thus be traced to redundant off-shell structures, not to the physics of the flow itself.

Here we apply the methods of the essential RG \cite{Baldazzi:2021ydj,Baldazzi:2021orb} to the problem of gauge dependence. As a continuation of \cite{Falls:2024noj}, we aim to clarify how gauge independence emerges once the flow is organised around its on-shell content, and to identify the corresponding gauge-independent sector within an essential renormalisation group scheme. To this end, we work with the Einstein--Hilbert action supplemented by a topological term and consider a proper-time flow in which the inessential couplings are fixed by field redefinitions. In dimensionless variables, the resulting flow equation separates naturally into the running of the cosmological sector, the Newton coupling, while the essential scheme is implemented by fixing the inessential couplings at their Gaussian values. We consider two approximations on top of the one-loop improved nature of the proper-time flow. 

First we use generak backgrounds to project onto the space of all terms with up to four derivatives exluding boundary terms. In this setup, the flow of the dimensionless Newton coupling can still exhibit an explicit dependence on the gauge parameter when its full off-shell form is retained. However, this dependence disappears when the flow is projected consistently onto the gauge-independent sector and expanded up to order $O(\tilde G^3)$. In this sense, the essential scheme already suggests that gauge dependence belongs to redundant directions in theory space and should drop out once the physical projection is performed~\cite{Falls:2024noj,Baldazzi:2021ydj,Baldazzi:2021orb,Knorr:2022ilz,Knorr:2023usb}.

Moving beyond four derivative terms, we shall also analyse the quadratic gauge-fixed action on a maximally symmetric background, keeping all powers in the Ricci scalar $R$. For a general class of two-parameter gauges, the fluctuation field is decomposed by means of the transverse-traceless split~\cite{Christensen:1979iy,Dou:1997fg,Reuter:2001ag}, and the resulting quadratic form is written in terms of tensor, vector, and scalar sectors. Off-shell, the corresponding one-loop effective action depends explicitly on the gauge parameters. This is reflected both in the structure of the Hessian and in the operators governing the non-physical sectors. The ghost determinant inherits the same gauge dependence, but away from the equations of motion it does not remove it completely. As a result, the full off-shell one-loop contribution is gauge dependent, in agreement with the general expectation that unphysical directions remain entangled with the flow before the on-shell projection is imposed.

Going on-shell, we observe simplifications: when the background satisfies the equations of motion, the gauge-dependent terms in the gauge-fixed action simplify drastically, and the ghost sector produces precisely the terms needed to cancel the residual gauge dependence. The complete one-loop contribution then becomes fully gauge independent. This cancellation is one of the central results of the paper: it shows explicitly that the gauge dependence of the proper-time flow is an off-shell effect, and that the physical one-loop content is recovered only after the correct on-shell combination of fluctuation and ghost sectors is taken into account. In this respect, the analysis provides a concrete implementation, within the proper-time framework~\cite{Bonanno:2004sy,Bonanno:2019ukb}, of the more general lesson of the on-shell approach~\cite{Benedetti:2011ct,Falls:2024noj,Falls:2017cze}: the physically meaningful part of the flow should be identified only after redundant gauge-dependent contributions, technically the BRST exact terms, have been removed.

The technical simplicity of proper-time flows, and their close connection to the structure of one-loop effective actions, is exactly what makes them a natural setting for examining how regularisation, gauge fixing, and on-shell projection interplay. The present construction indicates that the relevant question is not whether the off-shell flow is gauge independent in every intermediate step, but whether the flow can be organised so that all gauge dependence is confined to inessential sectors. Once this is achieved, the fixed-point structure extracted from the essential couplings should be understood as the physical content of the theory, while the remaining gauge dependence reflects the freedom to choose an off-shell representative.

From this point of view, the problem of gauge independence is part of a broader effort to formulate asymptotic safety in terms of essential and on-shell data. The on-shell perturbative construction of \cite{Falls:2024noj} shows that this can be done systematically in terms of parametrisation dependence. The analysis presented here suggests that the same logic applies to gauge dependence in the proper-time flow: the off-shell formulation retains gauge-dependent artifacts, but these disappear once the flow is projected onto its on-shell content and expressed in an essential scheme. This strengthens the interpretation of the proper-time flow as a useful tool for extracting universal information from quantum gravity, provided its results are organised around the physical sector from the outset.

The rest of this paper is organised as follows. In section~\ref{sec:setup}, we introduce the proper-time renormalisation group flow in the essential scheme and develop its curvature expansion in a one-loop improved approximation. Section~\ref{sec:truncatedflow} studies the four-derivative approximation, analysing the gauge dependence of the beta function for Newton's coupling and showing how gauge independence is recovered within the essential truncation. In section~\ref{sec:oneloop}, we examine the one-loop effective action on maximally symmetric backgrounds and demonstrate explicitly that the gauge-dependent contributions of the gauge-fixed action are cancelled on shell by the ghost sector. Section~\ref{sec:allorders} extends the essential scheme to all orders in the scalar curvature, showing how the gauge dependence can be shifted entirely into the field-redefinition kernel while the flow of the essential Newton coupling remains gauge independent. Finally, section~\ref{sec:conclusions} summarises our results and discusses their implications for asymptotically safe quantum gravity.

\section{Essential proper-time flow}
\label{sec:setup}

In this section, we introduce the proper-time renormalisation group framework used throughout the paper and explain how it is adapted to isolate the on-shell, gauge-independent sector of the flow. We implement the essential scheme idea in a proper-time renormalisation group flow. The flow is supplemented by a scale-dependent field redefinition of the metric, which removes components proportional to the equations of motion and thereby fixes the inessential directions. This provides the framework in which the gauge dependence of the off-shell flow can be separated from the running of the essential couplings.

\subsection{Essential scheme and generalised proper-time flow}

The central idea of the essential scheme is that not all couplings appearing in the running effective action should be treated on the same footing. Some correspond to physical directions in theory space, while others reflect the freedom to redefine the fields. In a Wilsonian language~\cite{Wegner:1974sla}, the latter are inessential couplings~\cite{Baldazzi:2021ydj,Baldazzi:2021orb,Knorr:2022ilz,Knorr:2023usb,Latorre:2000qc}. Their flow is not itself physical and can be fixed by an appropriate choice of renormalisation group coordinates. In the present context, this is implemented by supplementing the proper-time flow with a scale-dependent field redefinition of the metric.  We work with a proper-time renormalisation group equation whose right-hand side has a one-loop trace structure, but is evaluated with the scale-dependent Hessian of the running action. In this sense the flow resums contributions through the running couplings, rather than being an ordinary fixed one-loop perturbative calculation; while the one-loop order information is recovered we also include some diagrams from all orders in perturbation theory by the resummation. We use the background-field method as in \cite{Falls:2024noj}, splitting the metric according to the linear parametrization. Technically, we make two approximations. Firstly, we use the proper-time equation, which partially re-sums perturbation theory, but is expected to fail at two-loop order. Secondly, we truncate the effective action to a subset of all diffeomorphism-invariant terms. 

Accordingly, we consider the generalised proper-time flow equation for the scale-dependent effective action
\begin{equation}
\partial_t \Gamma_k
+
\int d^d x\,
\Psi_{\mu\nu}(x)\,
\frac{\delta \Gamma_k}{\delta g_{\mu\nu}(x)}
=
\Tr W(F)
-
\Tr W\!\left(\frac{1}{\alpha}Q\right)
-
\Tr W(Q),
\label{eq:generalised_PT_flow}
\end{equation}
with $t=\ln k/k_0$, where $k_0$ is a arbitrary reference scale, and $W$ the proper-time profile function. Here $\alpha$ is the gauge fixing parameter entering in
 the gauge-fixing term for the metric fluctuation $h_{\mu \nu}$
\begin{equation}
    S_{\text{gf}}=
    \frac{1}{16\pi G_N \alpha}
    \int d^d x\,
    \left(
        \nabla_\rho h^{\rho}{}_{\mu}
        -\frac{1+\beta}{d}\nabla_\mu h
    \right)
    \left(
        \nabla_\sigma h^{\sigma\mu}
        -\frac{1+\beta}{d}\nabla^\mu h
    \right)\,.
\end{equation}
Equation \eqref{eq:generalised_PT_flow} has the trace structure of a one-loop determinant, but it is used as a renormalisation group flow for the scale-dependent action $\Gamma_k$, improved by a scale-dependent field redefinition. The first trace represents the contribution of the gauge-fixed Hessian $F$, while the remaining two traces arise from the Faddeev--Popov sector. The additional term proportional to $\Psi_{\mu\nu}$ accounts for the freedom to perform $k$-dependent redefinitions of the metric field and is therefore proportional to the equations of motion. Its role is to remove the flow along redundant directions in theory space and to isolate an essential scheme in which the running of the physical couplings can be identified more cleanly. We note that we have chosen to include the gauge fixing parameter $\alpha$ in ``one half'' of the ghost contribution. As we will explain below, this is a choice of how the regulator depends on $\alpha$. The choice is crucial to ensure gauge-dependent terms cancel in the regularised theory as they do in a standard one-loop calculation, for example, when using dimensional regularisation.

The proper-time regulator is encoded in the profile function $W$, which we choose as
\begin{equation}
W(z)=\left(\frac{k^2}{z+k^2}\right)^m,
\label{eq:PT_profile}
\end{equation}
with $m$ the proper-time parameter. This choice suppresses the contribution of modes with $z\ll k^2$ and leaves the ultraviolet part of the spectrum essentially unaffected, thereby implementing the renormalisation group coarse graining in proper-time form. We do not discuss this construction in detail here, since both the proper-time regularisation~\cite{Bonanno:2004sy,Bonanno:2000yp,Mazza:2001bp,Bonanno:2019ukb,Abel:2023ieo,Glaviano:2024hie,Litim:2001ky,Wetterich:2024ivi,Bonanno:2025tfj,Bonanno:2025dry,Giacometti:2025qyy,Giacometti:2026zrs,Bonanno:2025qsc,Bonanno:2026mzs,Bonanno:2026aiw,Bonanno:2026ljx,Glaviano:2026lew, Bonanno:2026bbb} and the interpretation of the term proportional to the equations of motion~\cite{Baldazzi:2021ydj,Baldazzi:2021orb,Knorr:2023usb,Knorr:2022ilz,Falls:2024noj} are standard in the literature.

What does require a brief explanation is the structure of the ghost contribution on the right-hand side of \eqref{eq:generalised_PT_flow}. In the present gauge, the Faddeev--Popov determinant appears in the form
\begin{equation}
\det\!\left(\frac{Q}{\sqrt{\alpha}}\right).
\end{equation}
In the unregularised theory, one could of course keep this determinant as a single object. In the proper-time flow it is convenient to split it as
\begin{equation}
\det\!\left(\frac{Q}{\sqrt{\alpha}}\right)
=
\sqrt{\det\!\left(\frac{1}{\alpha}Q\right)}
\,
\sqrt{\det Q}.
\label{eq:ghost_split}
\end{equation}
Taking the logarithm, this leads precisely to the two ghost traces in
\eqref{eq:generalised_PT_flow},
\begin{equation}
-\Tr W\!\left(\frac{1}{\alpha}Q\right)-\Tr W(Q).
\end{equation}
The reason for this decomposition is that, once the theory is regularised, the cancellation between the gauge-dependent contributions of the gauge-fixed Hessian and those of the ghost sector is correctly implemented only if the two factors in \eqref{eq:ghost_split} are regularised separately. In this way, the gauge-parameter dependence is distributed between the two ghost contributions in exactly the form needed for the on-shell cancellation mechanism discussed below. For this reason, the split \eqref{eq:ghost_split} is an essential ingredient in the proper-time regularisation  if we want to achieve gauge independence. In particular, we recognise that in our approximation we have a regularisation dependence and that we have chosen a class of regulators which allow for the cancellation of the gauge dependence.

The identity~\eqref{eq:ghost_split} is algebraically natural: for any positive operator
$\mathcal Q$ and any positive gauge parameter $\alpha$,
\begin{equation}
\det\!\left(\frac{\mathcal Q}{\sqrt\alpha}\right)
\;=\;\alpha^{-N/2}\,\det\mathcal Q
\;=\;\sqrt{\det(\mathcal Q/\alpha)}\,\sqrt{\det \mathcal Q},
\end{equation}
with $N$ the dimension of the space on which $\mathcal Q$ acts. At the level of bare
determinants the two sides are equal and the rewriting carries no content. The non-trivial
point is that, once a regularisation is imposed, the proper-time profile $W$ does
\emph{not} respect this identity: $W$ acts as a function of the eigenvalues of its
argument. The choice adopted in~\eqref{eq:generalised_PT_flow} is not introduced by hand: it
reflects the mode-by-mode structure of the Faddeev--Popov determinant on the gauge-fixed
background. Decomposing the ghost vector field as $c^\mu = c^{T\mu}+\nabla^\mu c$, the
determinant factorises as
\begin{equation}
\det\mathcal M \;=\; \sqrt{\det\mathcal M'}\,\sqrt{\det\mathcal M''},
\label{eq:fp_mode_factorisation}
\end{equation}
the two factors collecting operators with and without gauge-parameter dependence in
their leading sectors. This decomposition is realised explicitly in
Section~\ref{sec:oneloop}: on a maximally symmetric background the ghost action factorises
as $S_{gh}=S'_{gh}+S''_{gh}$, see
\eqref{eq:ghost1new}--\eqref{eq:ghost2new}, with $S'_{gh}$ built from operators carrying
$1/\alpha$ in their leading sector
($\mathcal Q'_{TT}=\Delta_1/\alpha$,\,
$\hat{\mathcal Q}'=(2/\alpha d)\,(\tilde\Delta^\beta)^2/\tilde\Delta$)
and $S''_{gh}$ from operators that do not
($\mathcal Q''_{TT}=\Delta_1$,\,
$\hat{\mathcal Q}''=(2/d)\,\tilde\Delta$).

Although the identity is algebraic at the level of unregularised determinants, it becomes non-trivial in the proper-time flow because the profile $W$ acts on the eigenvalues of each operator separately. The split therefore specifies how the Faddeev–Popov modes are regularised. On maximally symmetric backgrounds this decomposition matches the transverse and longitudinal ghost operators to the corresponding non-physical sectors of the graviton Hessian, as shown explicitly in section \ref{sec:oneloop}. This mode-by-mode matching is what permits the on-shell cancellation of the gauge-dependent determinants.

\bigskip

 One realisation of an essential scheme, within the on-shell perturbative formulation of gravity, is the minimal essential scheme (MES): one fixes the running of each inessential coupling by means of a field redefinition, and then extracts the beta function of the essential coupling from the remaining, on-shell part of the flow. In pure gravity, the natural choice is to use the freedom associated with $\Psi_{\mu\nu}$ to fix the renormalisation of the cosmological coupling while removing also higher order terms which vanish when $R_{\mu\nu} = 0$. The scheme is therefore ``minimal''  because one imposes only the minimal renormalisation condition needed to remove the inessential coupling from the physical flow. This is precisely the logic underlying the MES reviewed in refs.~\cite{Baldazzi:2021ydj,Baldazzi:2021orb,Falls:2024noj,Knorr:2022ilz,Knorr:2023usb}, where field redefinitions are used to remove off-shell contributions and the running of Newton's coupling is extracted from an on-shell projection.

\subsection{Resolvent representation of the proper-time traces}
The profile function is chosen in proper-time form as in equation \eqref{eq:PT_profile}. Equation~\eqref{eq:generalised_PT_flow} differs from the standard proper-time flow by the term involving $\Psi_{\mu\nu}$. This extra term is proportional to the equations of motion and therefore vanishes on shell. Its role is to shift the renormalisation group trajectory along redundant directions, allowing one to remove the running of inessential couplings and to formulate the flow directly in terms of an essential scheme.

To evaluate the traces appearing on the right-hand side of \eqref{eq:generalised_PT_flow}, we use a resolvent representation of the fluctuation operator. Let $F$ denote the graviton Hessian, including the gauge-fixing contribution, and define the resolvent $G(c)$ by
\begin{equation}
(F+c)\,G(c)=1,
\qquad
T(c)=\Tr G(c).
\label{eq:resolvent_def}
\end{equation}
To ensure the evaluation of $T(c)$ is finite we evaluate it using dimensional regularisation i.e. we keep the spacetime dimension $d$ arbitary.
The proper-time trace can then be reconstructed from derivatives of $T(c)$. More precisely,
\begin{equation}
\Tr W(F)
=
\frac{c^m(-1)^{m-1}}{(m-1)!}
\frac{d^{\,m-1}}{dc^{\,m-1}}\,T(c)\Big|_{c=k^2}.
\label{eq:proper_time_from_resolvent}
\end{equation}
Once we have taken the derivatives to reach  $\Tr W(F)$, we can take $d\to 4$ for $m\geq 3$.
The same strategy applies to the gauge and ghost operators. In this way, the evaluation of the proper-time flow is reduced to the computation of the integrated resolvent and its derivatives.

The advantage of this formulation is that it interfaces naturally with a curvature expansion. One writes the inverse operator in the form
\begin{equation}
G(c)=G_0(c)\,(1+M(c))^{-1},
\label{eq:G_split}
\end{equation}
where $G_0(c)$ is the flat-space inverse propagator and $M(c)$ contains the curvature-dependent corrections. Expanding in powers of $M(c)$ then generates the proper-time trace order by order in curvature~\cite{Barvinsky:1985an,Barvinsky:1990up,Barvinsky:2021ijq,Barvinsky:2024kgt,Groh:2011dw,Ferrero:2023xsf,Bastianelli:2013tsa,Bastianelli:2022pqq}. This is the sense in which the present analysis remains structurally close to the one-loop determinant: the flow is represented through the same differential operators that govern the Gaussian integration over fluctuations, but is organised in a way that permits a direct projection onto essential couplings.

Operationally, this expansion gives access to the coefficients of the local invariants appearing in the truncation. The coefficient of the volume term determines the flow of the cosmological sector, the coefficient of $R$ determines the running of Newton's coupling, and the coefficient of $\Xi$ (introduced below) controls the topological curvature-squared sector. Since our purpose is to identify the gauge-independent essential flow, the crucial step is to separate from these coefficients all contributions proportional to the equations of motion, which can be reabsorbed into the field redefinition kernel.

\subsection{Curvature-squared truncation}
We consider Euclidean gravity in $d$ dimensions and work with  the scale dependent effective  action in the Einstein--Hilbert action supplemented by the Euler density,
\begin{equation}
\Gamma_k[g]
=
\int d^d x \sqrt{g}\,
\left(
\frac{\rho_k}{8\pi}
-\frac{R}{16\pi G_k}
+\vartheta_k\, \Xi
\right),
\label{eq:Gamma_ansatz}
\end{equation}
where
\begin{equation}
\Xi = R^2 - 4 R_{\mu\nu}R^{\mu\nu} + R_{\mu\nu\rho\sigma}R^{\mu\nu\rho\sigma}.
\label{eq:Euler_density}
\end{equation}
Here $k$ is the RG scale, $G_k$ is Newton's coupling, $\rho_k$ denotes the cosmological coupling, and $\vartheta_k$ multiplies the curvature-squared Euler density. In four dimensions the integral of $\Xi$ is topological and therefore does not contribute to the local equations of motion. It is retained since it parametrises the curvature-squared part of the truncation and naturally appears in the curvature expansion of the flow. While a term proportional to $R^2$  and to $R_{\mu\nu}R^{\mu\nu}$ could be added to the action, we use field redefinitions to set its coefficient to zero.

We now turn our attention to the on-shell projection.
The equations of motion following from \eqref{eq:Gamma_ansatz} are
\begin{equation}
\frac{\delta \Gamma_k}{\delta g_{\mu\nu}}
=
\frac{\sqrt{g}}{16\pi G_k}
\left(
R_{\mu\nu}
-\frac{1}{2}R g_{\mu\nu}
+\Lambda_k g_{\mu\nu}
\right),
\qquad
\Lambda_k \equiv G_k \rho_k .
\label{eq:EOM}
\end{equation}
This equation makes the logic of the essential scheme transparent. Any contribution to the flow proportional to
\begin{equation}
R_{\mu\nu}
-\frac{1}{2}R g_{\mu\nu}
+\Lambda_k g_{\mu\nu}
\label{eq:EOM_tensor}
\end{equation}
is redundant, since it can be absorbed into the field redefinition $\Psi_{\mu \nu}$.
In the truncation considered in this section, we take the field-redefinition kernel to admit a curvature expansion of the form
\begin{equation}
\Psi_{\mu\nu}
=
\gamma_g\, g_{\mu\nu}
+
\gamma_R\, R\, g_{\mu\nu}
+
\gamma_{Ricci}\, R_{\mu\nu},
\label{eq:Psi_expansion}
\end{equation}
where $\gamma_g$, $\gamma_R$, and $\gamma_{Ric}$ are scale-dependent coefficients. These are not new physical couplings. Rather, they parametrise the freedom to make field transformations and will be fixed by the requirement that the flow be projected onto its essential sector. In the present work, we use this freedom to fix the inessential cosmological coupling $\rho_k$, so that the physically relevant output of the flow is encoded in the running of the remaining essential couplings.
In other words, the generalised flow \eqref{eq:generalised_PT_flow} should not be interpreted as assigning physical significance to the full off-shell running of $\Gamma_k$. Its purpose is rather to disentangle the essential part of the flow from the directions that vanish on shell.

This point is important for the interpretation of gauge dependence. Away from the equations of motion, the flow generically depends on the gauge-fixing parameters and on the choice of fluctuation variables. Such dependence is not surprising because off-shell quantities are not observables. The relevant question is whether the running extracted after projection onto the on-shell sector remains gauge dependent. The working hypothesis of the present construction is that, once the flow is organised in an essential scheme, gauge dependence can be confined to redundant structures, while the beta functions of the essential couplings become gauge independent, at least within the truncations considered here.

For later convenience, it is useful to introduce the shorthand
\begin{equation}
-d\, \Omega_{\mu\nu}
\equiv
R_{\mu\nu}
-\frac{1}{2}R g_{\mu\nu}
+\Lambda_k g_{\mu\nu},
\label{eq:Omega_tensor}
\end{equation}
so that the equations of motion read simply $\Omega_{\mu\nu}=0$. The distinction between off-shell and on-shell quantities will play a central role in the later sections. In particular, the one-loop analysis will show explicitly that the full off-shell expression remains gauge dependent, whereas the complete on-shell contribution becomes gauge independent once the ghost sector is included.

The flow equation \eqref{eq:generalised_PT_flow} is evaluated in a local curvature expansion.
More precisely, both the effective action and the traces on the right-hand side are expanded in local
invariants built from the curvature of the background metric. In principle, such an expansion generates
not only powers of the algebraic invariants
but also terms containing derivatives of the curvature.
Using the field redefinitions, we project onto the subspace spanned by the Einstein-Hilbert action
together with the corresponding redundant structures generated by the field redefinition kernel
\eqref{eq:Psi_expansion}.

It is convenient to rewrite the theory in terms of dimensionless variables. Denoting by $\tilde G$,
$\tilde\rho$, and $\tilde\vartheta$ the dimensionless counterparts of $G_k$, $\rho_k$, and
$\vartheta_k$, the projected flow separates naturally into the cosmological sector, the Newton coupling,
and the curvature-squared contribution. The MES is then implemented by imposing a
renormalisation condition on the inessential coupling $\tilde\rho$, which we fix at its Gaussian value.
In this way, the coefficients $\gamma_g$, $\gamma_R$, and $\gamma_{\Ric}$ are determined by the
requirement that the flow has no component along the corresponding redundant directions.

The generalised proper-time equation is therefore used to choose the field-redefinition term so that the flow along redundant directions is fixed; the remaining coefficients then give the beta functions of the essential couplings. Within this interpretation, the
remaining beta functions encode the physically meaningful output of the truncation, while any gauge
dependence that survives only in the field-redefinition kernel should be regarded as inessential.

In later sections, we will sometimes specialise the discussion to maximally symmetric backgrounds, for which
the fluctuation operators simplify and the decomposition into tensor, vector, and scalar sectors becomes
particularly transparent. However, the conceptual structure of the essential scheme does not rely on that
restriction. The role of the background choice is only to provide a convenient arena in which the
cancellation of gauge-dependent terms can be displayed explicitly. The underlying distinction between
essential on-shell data and redundant off-shell contributions is more general.

The rest of the paper develops the consequences of this setup in two complementary directions. First, we
analyse the truncated proper-time flow of the essential couplings and identify the sector in which the beta
function for Newton's coupling becomes gauge independent. Second, we study the one-loop effective action in
more detail and show how the gauge-dependent off-shell terms cancel once the flow is projected on shell.
Together, these two analyses support the interpretation that the physically meaningful content of the
proper-time flow is encoded in its essential, on-shell sector.

\section{Four derivative approximation}
\label{sec:truncatedflow}

Having set up the proper-time flow in an essential scheme, we now turn to its truncated realisation in the local curvature expansion. The purpose of this section is twofold. First, we derive the beta functions associated with the Einstein--Hilbert truncation supplemented by the Euler density, working within the curvature-expanded ansatz introduced in the previous section. In particular we retain all terms up to order four derivatives (neglecting boundary terms). Second, we show that although the full off-shell beta function of Newton's coupling retains an explicit gauge dependence, this dependence disappears once the flow is projected consistently onto the essential on-shell sector and truncated at the appropriate order in $G$, namely to order $G^3$. In this way, the truncated proper-time flow already exhibits the central mechanism that will later be demonstrated by an explicit one-loop calculation. In this section, we put $\beta=1$, retaining the dependence only on $\alpha$.

We begin from the truncation \eqref{eq:Gamma_ansatz} and from the generalised proper-time flow described in section~\ref{sec:setup}. Since the only term with four derivatives of the metric with a derivative of the curvature is a total derivative (i.e. the integral of $\nabla^2 R$), we can neglect derivatives of curvature.

In our proper-time implementation, this idea is realised in the MES by allowing the kernel
$
\Psi_{\mu\nu}$
to absorb those components of the flow that are proportional to the equations of motion. The coefficients $\gamma_g, \gamma_R$, and $\gamma_{\Ric}$ are not new physical couplings nor beta functions. 

Introducing the dimensionless quantities
\begin{equation}
\tilde G = k^{d-2} G_k,
\qquad
\tilde \rho = k^{-d}\rho_k,
\qquad
\tilde \vartheta = k^{4-d}\vartheta_k,
\label{eq:dimlesscouplings_sec3}
\end{equation}
together with
\begin{equation}
\tilde R = k^{-2}R,
\qquad
\tilde R_{\mu\nu}=k^{-2}R_{\mu\nu},
\qquad
\tilde \Xi = k^{-4}\Xi,
\qquad
\tilde\gamma_R = k^2\gamma_R,
\qquad
\tilde\gamma_{\Ric}=k^2\gamma_{\Ric},
\label{eq:dimlesscurvatures_sec3}
\end{equation}
the left-hand side of the flow equation can be organised according to the independent invariants appearing in the truncation. Matching terms order by order in curvature yields
\begin{align}
\int d^d x \sqrt{g}\, k^d \Bigg\{
&
\tilde \Xi \Big[(d-4)\tilde\vartheta+\partial_t\tilde\vartheta\Big]
+
\frac{\tilde R_{\mu\nu}\tilde R^{\mu\nu}}{16\pi \tilde G}\,\tilde\gamma_{\Ric}
-
\frac{\tilde R^2}{32\pi \tilde G}\,
\Big[(d-2)\tilde\gamma_R+\tilde\gamma_{\Ric}\Big]
\nonumber\\[0.3em]
&
\hspace{2cm}
+
\frac{\tilde R}{32\pi \tilde G^2}
\Big[
2\,\partial_t \tilde G
-
(d-2)(2+\gamma_g)\tilde G
+
2(d\tilde\gamma_R+\tilde\gamma_{\Ric})\tilde\rho\,\tilde G^2
\Big]
\nonumber\\[0.3em]
&
\hspace{2cm}
+
\frac{1}{16\pi}
\Big[
2\,\partial_t \tilde\rho
+
d\,\tilde\rho\,(2+\gamma_g)
\Big]
\Bigg\}
=
\Tr W(F)-\Tr W\!\left(\frac1\alpha Q\right)-\Tr W(Q).
\label{eq:dimlessflowmatching_sec3}
\end{align}
To compute the right-hand side of this equation, we use eq. \eqref{eq:proper_time_from_resolvent}. The corresponding $G_0(c)$, $M(c)$ and $T(c)$, including the ghost contributions, are provided in the ancillary .nb file.
This expression makes transparent how the flow decomposes into the cosmological sector, the Einstein--Hilbert sector, and the curvature-squared sector. In particular, the coefficients of the volume term, of $\tilde R$, and of $\tilde \Xi$ determine $\partial_t\tilde\rho$, $\partial_t\tilde G$, and $\partial_t\tilde\vartheta$, respectively, while the coefficients of $\tilde R^2$ and $\tilde R_{\mu\nu}\tilde R^{\mu\nu}$ determine $\tilde\gamma_R$ and $\tilde\gamma_{\Ric}$.

\begin{figure}[h]
    \centering
   \includegraphics[width=.5\linewidth]{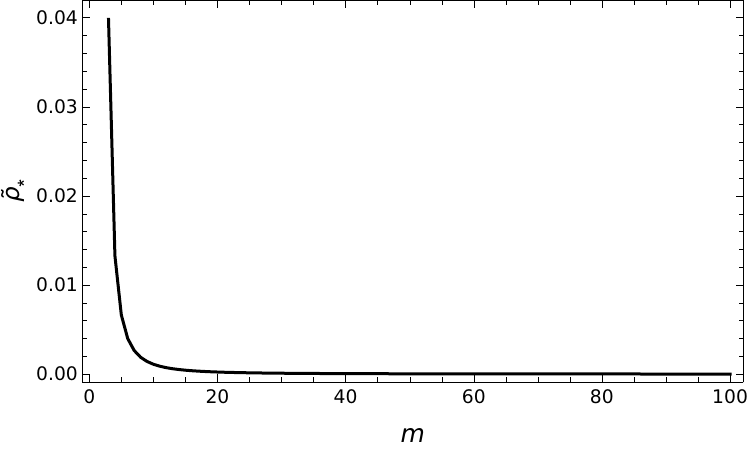}
    \caption{The inessential coupling $\tilde{\rho}_*$ in $d=4$ dimensions at the non-Gaussian fixed point for different values of $m$.}
    \label{inessential coupling}
\end{figure}

The MES is then implemented by fixing the inessential coupling $\tilde\rho$ rather than allowing it to run freely. Concretely, we impose
\begin{equation}
\partial_t \tilde\rho = 0,
\label{eq:rhofixed_sec3}
\end{equation}
and use this condition to determine $\gamma_g$. Furthermore, $\tilde \rho$ is fixed to its value at the Gaussian fixed point where $\tilde G =0$, this value is plotted in fig~\ref{inessential coupling}. This is the defining renormalisation condition of the scheme. Once it is imposed, the cosmological coupling is no longer part of the physical running, and the beta function for $\tilde G$ is extracted from the remaining projected flow. The coefficients $\gamma_R$ and $\gamma_{\Ric}$ then encode the residual off-shell reparametrisation freedom associated with the curvature-squared sector. Thus the content of the MES is not that all redundant structures are eliminated once and for all, but that the inessential coupling is fixed and the essential beta function is obtained only after this projection has been performed. 

\subsection{Gauge (in)dependence}

The resulting flow of $\tilde G$ depends in general on the gauge fixing. This is already visible at the level of the traces on the right-hand side of \eqref{eq:dimlessflowmatching_sec3}, since the graviton, gauge, and ghost operators are all gauge dependent away from the equations of motion. Consequently, the coefficient of $\tilde R$ in the full off-shell flow yields a beta function
\begin{equation}
\beta_{\tilde G} \equiv \partial_t \tilde G
=
\beta_{\tilde G}^{\rm off}(\tilde G,\tilde\rho;\alpha,m),
\label{eq:betaGoff_sec3}
\end{equation}
which depends explicitly on the gauge parameter $\alpha$, as well as on the proper-time regulator parameter $m$.

This feature reflects the fact that the coefficient of $R$ extracted from the  off-shell flow still contains contributions from redundant directions. In the present scheme these are precisely the contributions that are tied to the field redefinition kernel and to the off-shell continuation of the action. The point of the essential projection is to isolate the part of $\beta_{\tilde G}$ that survives once these redundant components have been removed.

To make this explicit, we expand the beta function in powers of $\tilde G$. If we were to use the standard proper-time equation with $\Psi_{\mu\nu}=0$ we would obtain 
\begin{equation}
\beta_{\tilde G}
=
(d-2)\tilde G
-
B_1(\alpha,m)\,\tilde G^2
-
B_2(\alpha,m,\tilde\rho)\,\tilde G^3
+
O(\tilde G^4),
\label{eq:betaGexpand_sec3}
\end{equation}
where the coefficients $B_1$ and $B_2$ are determined by the curvature expansion of the proper-time traces. In the full off-shell expression, these coefficients depend on the gauge parameter.\\ 
 
 The picture changes if the flow is consistently projected onto the essential on-shell sector using $\Psi_{\mu\nu}$ to implement the MES. Then the gauge-dependent contributions cancel up to the order $\tilde G^3$. In particular the truncated beta function reduces to a gauge-independent expression of the form
\begin{equation}
\beta_{\tilde G}^{\rm ess}
=
(d-2)\tilde G
-
\bar B_1(m)\,\tilde G^2
-
\bar B_2(m)\,\tilde G^3
+
O(\tilde G^4)\,,
\label{eq:betaGess_sec3}
\end{equation}
where up to order $\tilde G^3$ there is no dependence on the gauge.
Thus, within the present truncation, gauge independence is recovered not for the off-shell flow with $\Psi_{\mu\nu} = 0$, but for its essential projection via the MES. Nonetheless, the full beta function for all orders in $\tilde G$ retains a strong dependence on the gauge.  

This result is important and guides our approach. It shows that the gauge dependence of the off-shell beta function is not an intrinsic ambiguity in the physical running of Newton's coupling, but rather a signal that the flow has not yet been stripped of redundant information within a given approximation. The essential scheme therefore does more than merely reorganise the computation: it identifies the sector of the flow that can plausibly be assigned universal meaning. While physical information computed exactly would be independent of the gauge, in any scheme, the essential scheme allows one to realise gauge independence in a relatively simple approximation. 

\begin{figure}[h]
    \centering
    \includegraphics[width=1\linewidth]{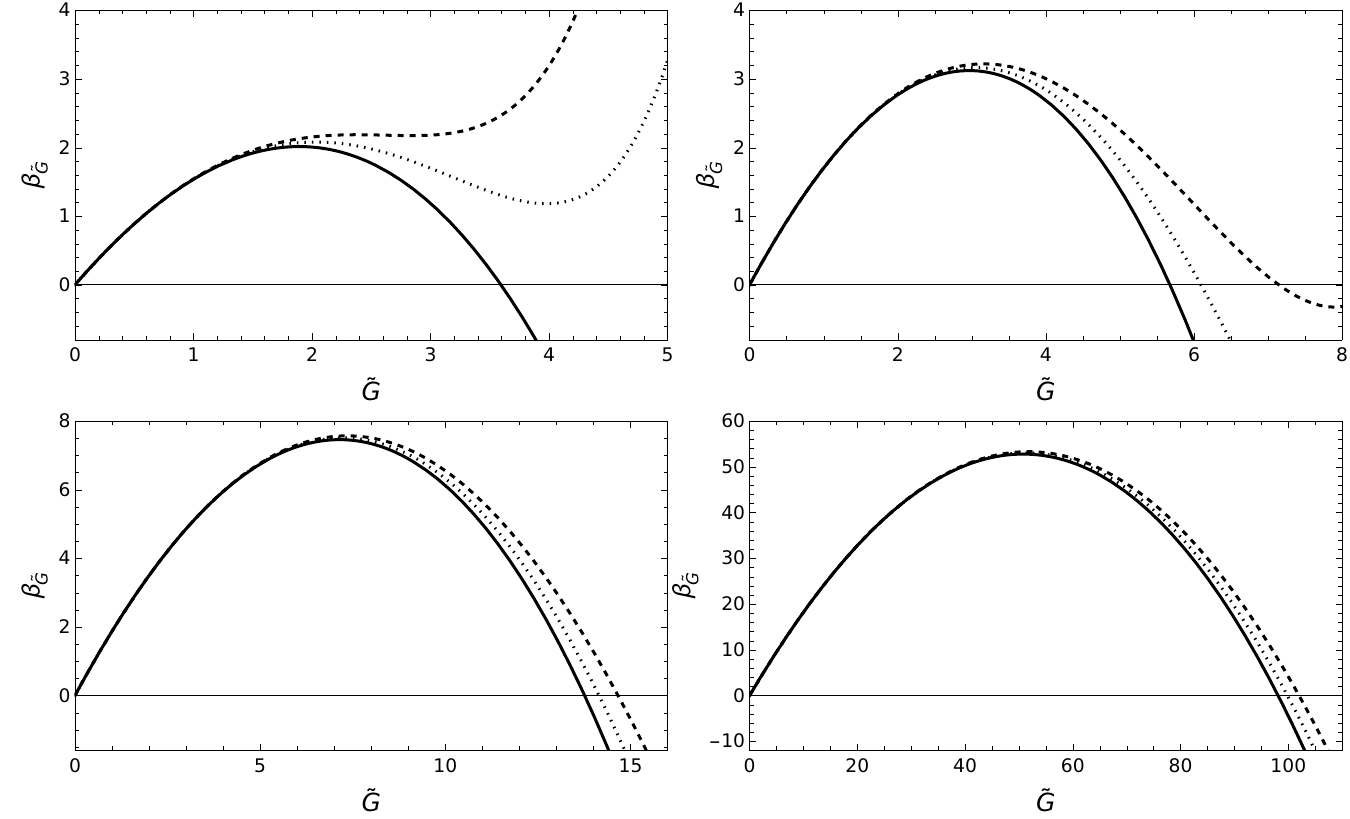}
    \caption{Beta function of the dimensionless Newton constant in $d=4$ dimensions for different scenarios: $m=3$ (upper left), $m=4$ (upper right), $m=8$ (bottom left), $m=50$ (bottom right). In each plot the solid line shows the gauge independent truncation up to $O(\tilde{G}^3)$, the dotted line the full beta function in the Feynman gauge and the dashed line the one in the Landau gauge.}
    \label{beta functions plot}
\end{figure}

\subsection{Fixed point structure and critical exponents}

We now turn to the fixed-point analysis implied by the  essential flow. The inessential coupling $\tilde\rho$ is fixed by construction, and the resulting value $\tilde\rho_\ast$ is gauge independent within the present approximation. As shown in Figure \ref{inessential coupling}, however, it retains a dependence on the proper-time regulator parameter $m$.

The main object of interest is the non-trivial zero of the beta function \eqref{eq:betaGess_sec3},
\begin{equation}
\beta_{\tilde G}^{\rm ess}(\tilde G_\ast)=0,
\label{eq:NGFP_sec3}
\end{equation}
which defines the non-Gaussian fixed point of the Newton coupling. In $d=4$, and for a broad range of values of the proper-time parameter $m$, the essential beta function exhibits a positive fixed point
\begin{equation}
\tilde G_\ast > 0,
\label{eq:Gstarpositive_sec3}
\end{equation}
which is ultraviolet attractive.

More specifically, Figure \ref{beta functions plot} shows that, for $m=3$, the beta function exhibits a strong gauge dependence, such that the non-Gaussian fixed point disappears once the $O(\tilde{G}^4)$ terms are included. As $m$ increases, the impact of these higher order terms, encoding the gauge dependence, becomes progressively weaker.

\begin{figure}[h]
    \centering
   \includegraphics[width=.5\linewidth]{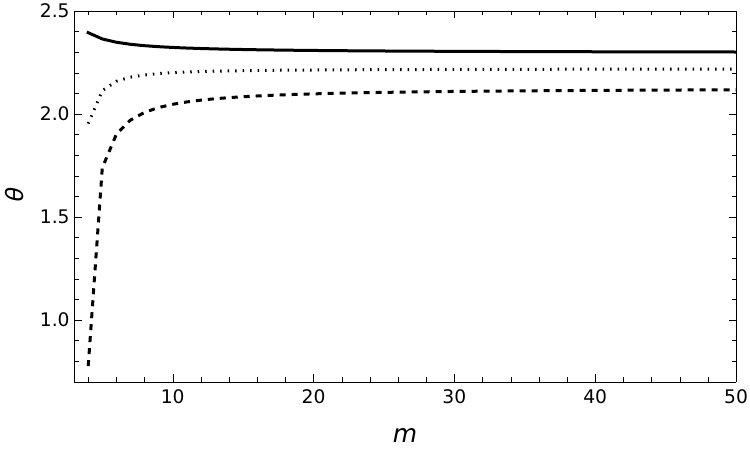}
    \caption{The critical exponent $\theta$ in $d=4$ dimensions at the non-Gaussian fixed point for different values of $m$. The solid line shows the result for the gauge independent truncation up to $O(\tilde{G}^3)$, the dotted line for the full beta function in the Feynman gauge $\alpha =1$ and the dashed line the one in the Landau gauge $\alpha =0$.}
    \label{critical exponent plot}
\end{figure}

The corresponding critical exponent is defined by
\begin{equation}
\theta
=
-
\left.
\frac{\partial \beta_{\tilde G}^{\rm ess}}{\partial \tilde G}
\right|_{\tilde G=\tilde G_\ast},
\label{eq:criticalexp_sec3}
\end{equation}
and is found to be positive throughout the same range of regulator values. Figure \ref{critical exponent plot} shows that the gauge dependence becomes progressively weaker as $m$ increases, although a residual dependence remains.

The existence of this fixed point and the positivity of $\theta$ indicate that the UV-attractive structure familiar from proper-time flows survives the essential projection.

This observation is one of the central results of the truncated analysis. Even though the full off-shell flow depends on the gauge fixing, the non-Gaussian fixed point persists in the gauge-independent essential sector. In this sense, the fixed-point behaviour found here is not merely an artifact of a particular off-shell representation, but survives the projection onto the part of the flow that is expected to encode its physical content. This is consistent with previous proper-time and effective-average-action analyses in the Einstein--Hilbert truncation and its extensions~\cite{Reuter:1996cp,Reuter:2001ag,Bonanno:2004sy,Codello:2007bd,Machado:2007ea,Falls:2014zba,Falls:2015qga,Falls:2016msz,Baldazzi:2021orb,Falls:2024noj}.

Finally, the flow of the topological coupling is particularly simple. In $d=4$, the leading term in the beta function of $\tilde\vartheta$ is independent of both the gauge parameter and the regulator parameter, and takes the form
\begin{equation}
\partial_t \tilde\vartheta
=
\frac{53}{720\pi^2}
+
O(\tilde G).
\label{eq:betatheta_sec3}
\end{equation}
This serves as an additional consistency check on the curvature expansion, since the coefficient of the Euler density is insensitive to the gauge-fixing parameter already at leading order. Starting at order $\tilde G$, the beta function for $\tilde\vartheta$ depends on the gauge.

 From the truncated proper-time flow, we observe the following. The off-shell beta function of Newton's coupling is gauge dependent, as expected from a flow equation formulated for a gauge-fixed effective action away from the equations of motion. Once the flow is projected onto the essential sector and expanded consistently in powers of $\tilde G$, the gauge dependence disappears to the order considered. The resulting beta function exhibits a UV-attractive non-Gaussian fixed point with positive critical exponent.

This conclusion should be interpreted with due caution. The present analysis is performed within a restricted truncation and within a proper-time framework that retains a one-loop-like structure. It therefore does not establish complete gauge independence of the full non-perturbative theory. What it does show is that the apparent gauge ambiguity of the off-shell flow can be removed, at least in this setting, by isolating the essential on-shell sector. This provides strong evidence that the physical content of the proper-time flow lives in this sector. As with the dependence on the parametrisation \cite{Falls:2024noj}, gauge independence is found when we expand the beta function to cubic order in Newton's constant, in our truncation up to four orders in derivatives. Higher orders in $G$ are dependent on the gauge.

\section{Explicit gauge independence at one loop}
\label{sec:oneloop}

The analysis of the previous section was already based on a generalized proper-time
flow with the usual one-loop trace structure. The purpose of the present section
is therefore not to introduce a different approximation, but to make explicit,
at the level of the one-loop effective action, the on-shell cancellation mechanism
underlying the gauge-independent sector identified there.

Although the recovery of gauge independence is not, in principle, restricted to
maximally symmetric backgrounds, such backgrounds provide a particularly convenient
arena in which the fluctuation operators simplify and the cancellation can be
displayed in closed form. We therefore use them here as an explicit realisation
of the general on-shell mechanism. Here we retain dependence on both $\alpha$ and $\beta$.

We specialise to a maximally symmetric background,
\begin{equation}
    R_{\mu\nu\rho\sigma}
    =
    \frac{R}{d(d-1)}
    \left(
        g_{\mu\rho}g_{\nu\sigma}
        -g_{\mu\sigma}g_{\nu\rho}
    \right),
    \qquad
    R_{\mu\nu}
    =
    \frac{R}{d}g_{\mu\nu}.
\end{equation}
We decompose the fluctuation field according to the TT split~\cite{Christensen:1979iy,Dou:1997fg,Reuter:2001ag,Codello:2007bd},
\begin{equation}
    h_{\mu\nu}
    =
    h^{TT}_{\mu\nu}
    +\nabla_\mu \xi_\nu
    +\nabla_\nu \xi_\mu
    +\left(
        \nabla_\mu\nabla_\nu
        +\frac{1}{d}g_{\mu\nu}\Delta
    \right)\sigma
    +\frac{1}{d}g_{\mu\nu}h.
\end{equation}
The quadratic part of the gauge-fixed action then takes the form
\begin{dmath}\label{eq:hessianTTnew}
\delta^2\!\left(\Gamma_k+S_{\text{gf}}\right)
=
\frac{1}{32\pi G_N}
\int d^d x \sqrt{g}\,
\Bigg\{
h^T_{\mu\nu}\left(\Delta_2+2\Omega\right)h^{T\,\mu\nu}
+\xi_\mu\left(\frac{2}{\alpha}\left(\Delta_1\right)^2+4\Omega\Delta_1\right)\xi^\mu
-\sigma\Delta\tilde{\Delta}\frac{2}{d^2}
\left[
\left(
\frac{d-2}{2}+\frac{1-d}{\alpha}
\right)\Delta
+\frac{R}{\alpha}
-d\Omega
\right]\sigma
-2\sigma
\frac{\alpha(d-2)-2\beta}{\alpha d^2}
\Delta\tilde{\Delta}\,h
+h\left[
\tilde{\Delta}^{\alpha\beta}
-\frac{d-2}{d}\Omega
\right]h
\Bigg\},
\end{dmath}
where
\begin{equation}
    \Delta_1=\Delta-\frac{R}{d},
    \qquad
    \Delta_2=\Delta+\frac{2R}{d(d-1)},
    \qquad
    \tilde{\Delta}=(d-1)\Delta-R,
\end{equation}
\begin{equation}
    \tilde{\Delta}^{\alpha\beta}
    =
    \frac{1}{d^2}
    \left(
        2\frac{\beta^2}{\alpha}\Delta-(d-2)\tilde{\Delta}
    \right),
    \qquad
    \Omega
    =
    R\frac{d-2}{2d}-\Lambda.
\end{equation}

The one-loop effective action is obtained by combining the Gaussian integration
over the fluctuation fields with the Faddeev--Popov ghost determinant. Off shell,
both quantities depend explicitly on the gauge parameters $\alpha$ and $\beta$.
The physically relevant question is whether this gauge dependence survives after
projection onto the equations of motion.

The on-shell condition reads
\begin{equation}
    \Omega=0\,.
\end{equation}
On shell, the gauge-dependent contributions in the scalar and vector sectors
simplify drastically. The ghost sector then produces the counterterms required to
cancel the remaining gauge dependence of the gauge-fixed action. The complete
one-loop contribution becomes gauge independent only after this full combination
of fluctuation and ghost sectors is taken into account.

\subsection{Explicit realisation on maximally symmetric backgrounds}

Each field in the TT decomposition is expanded in orthogonal Laplacian eigenmodes,
\begin{equation}
    h^{TT}_{\mu\nu}
    =
    \sum_{n=2}^{\infty}\sum_{m=1}^{D_n(d,2)}
    h^{TT}_{nm}T^{nm}_{\mu\nu},
\end{equation}
\begin{equation}
    \xi_\mu
    =
    \sum_{n=1}^{\infty}\sum_{m=1}^{D_n(d,1)}
    \xi_{nm}T^{nm}_{\mu},
\end{equation}
\begin{equation}
    \sigma
    =
    \sum_{n=0}^{\infty}\sum_{m=1}^{D_n(d,0)}
    \sigma_{nm}T^{nm},
\end{equation}
\begin{equation}
    h
    =
    \sum_{n=0}^{\infty}\sum_{m=1}^{D_n(d,0)}
    h_{nm}T^{nm},
\end{equation}
with
\begin{equation}
    \Delta T^{nm}_{\mu\nu}=\lambda_n^2 T^{nm}_{\mu\nu},
    \qquad
    \Delta T^{nm}_{\mu}=\lambda_n^1 T^{nm}_{\mu},
    \qquad
    \Delta T^{nm}=\lambda_n^0 T^{nm},
\end{equation}
and
\begin{equation}
    \lambda_n^s
    =
    \frac{R}{d(d-1)}
    \left(
        n(n-d-1)-s
    \right).
\end{equation}
The corresponding multiplicities~\cite{Kluth:2019vkg,Groh:2011dw,Benedetti:2010nr} are
\begin{equation}
    D_n(d,0)
    =
    \frac{(n+d-2)!(2n+d-1)}{n!(d-1)!},
    \qquad n=0,1,\dots,
\end{equation}
\begin{equation}
    D_n(d,1)
    =
    \frac{n(n+d-1)(n+d-3)!(2n+d-1)}{(n+1)!(d-2)!},
    \qquad n=1,2,\dots,
\end{equation}
\begin{equation}
    D_n(d,2)
    =
    \frac{(d+1)(d-2)(n+d)(n-1)(n+d+3)!(2n+d-1)}
    {2(n+1)!(d-1)!},
    \qquad n=1,2,\dots.
\end{equation}

The Gaussian measure induced by the metric on field space can be written as
\begin{align}
    \nonumber
    &\int d^dx\,d^dy\,
    \gamma^{\mu\nu,\rho\sigma}(x,y)h_{\mu\nu}(x)h_{\rho\sigma}(y)
    \\
    &=
    \frac{1}{32\pi G}
    \int d^d x \sqrt{g}
    \left[
        h^{TT}_{\mu\nu}h^{TT\,\mu\nu}
        +\xi_\mu(2\Delta_1)\xi^\mu
        +\sigma\left(\frac{1}{d}\Delta\tilde{\Delta}\right)\sigma
        +h\left(\frac{2-d}{2d}\right)h
    \right],
\end{align}
which in terms of the mode coefficients becomes
\begin{align}
    \nonumber
    &=
    \frac{1}{32\pi G_N}
    \Bigg[
    \sum_{n=2}^{\infty}\sum_{m=1}^{D_n(d,2)}
    \left(h^{TT}_{nm}\right)^2
    +
    \sum_{n=2}^{\infty}\sum_{m=1}^{D_n(d,1)}
    2\lambda'_n(\xi_{nm})^2
    \\
    &\qquad\qquad
    +
    \sum_{n=2}^{\infty}\sum_{m=1}^{D_n(d,0)}
    \frac{\lambda_n^0\tilde{\lambda}_n}{d}(\sigma_{nm})^2
    +
    \sum_{n=0}^{\infty}\sum_{m=1}^{D_n(d,0)}
    \left(\frac{2-d}{2d}\right)(h_{nm})^2
    \Bigg],
\end{align}
where
\begin{equation}
    \lambda'_n=\lambda_n^1-\frac{R}{d},
    \qquad
    \tilde{\lambda}_n=(d-1)\lambda_n^0-R.
\end{equation}
The modes $\lambda'_1=\lambda_0=\tilde{\lambda}_1=0$ do not contribute to
the sum. The conformal sector has the wrong sign, and following the Gibbons--Hawking--Perry
prescription~\cite{Gibbons:1978ac} we rotate $h\to ih$.\footnote{
We use the Gibbons–Hawking–Perry prescription as a practical recipe for rendering the conformal-mode integral convergent. Recent works have reinterpreted the closely related $i^{D+2}$ phase factor of the Euclidean gravitational partition function as an observer-dependent artifact, arguing that it is largely cancelled once a physical observer carrying the appropriate negative modes is included in the path integral \cite{Maldacena:2024spf, Ivo:2025yek}. A detailed analysis of this interpretation, and of its possible implications for proper-time flows, lies beyond the scope of the present work, where the conformal rotation enters only as a regularisation step prior to the on-shell projection.} The functional measure then becomes
\begin{equation}
    Dh_{\mu\nu}
    =
    J\,Dh^{TT}_{\mu\nu}D\xi_\mu D\sigma Dh,
\end{equation}
with Jacobian
\begin{align}
    \nonumber
    J=&
    \left(
        \prod_{n=2}^{\infty}\prod_{m=1}^{D_n(d,2)}
        \frac{1}{\sqrt{32\pi G_N}}
    \right)
    \left(
        \prod_{n=2}^{\infty}\prod_{m=1}^{D_n(d,1)}
        \sqrt{\frac{\lambda'_n}{16\pi G_N}}
    \right)
    \\
    &\times
    \left(
        \prod_{n=2}^{\infty}\prod_{m=1}^{D_n(d,0)}
        \sqrt{\frac{\lambda_n^0\tilde{\lambda}_n}{32\pi G_N d}}
    \right)
    \left(
        \prod_{n=0}^{\infty}\prod_{m=1}^{D_n(d,0)}
        \sqrt{\frac{d-2}{64\pi G_N d}}
    \right).
\end{align}
It is convenient to introduce a second scalar decomposition
\begin{equation}
    \tilde{h}:=\sum_{n=0}^1 \sum_{m=1}^{D_n(d,0)} h_{nm}T^{lm},
\end{equation}
and
\begin{equation}
    \hat{h}:=h-\tilde{h}=\sum_{n=2}^\infty \sum_{m=1}^{D_n(d,0)} h_{nm}T^{lm}.
\end{equation}
Then
\begin{equation}
    \bar{h}=\hat{h}+\Delta \sigma,
\end{equation}
and
\begin{equation}
    \bar{\sigma}=\sigma+\frac{\beta}{(d-1-\beta)\Delta-R}\bar{h}.
\end{equation}
On shell, after this second scalar decomposition, the
quadratic form diagonalises to
\begin{align}
    \nonumber
    \delta^2_{\text{OnShell}}(S+S_{\text{gf}})
    =
    \int d^dx \sqrt{g}\Bigg[
    h^{TT}_{\mu\nu}\Delta''_2 h^{TT\,\mu\nu}
    +\xi_\mu\frac{1}{\alpha}(\Delta_1)^2\xi^\mu
    +\bar{\sigma}\frac{2}{\alpha d^2}\Delta(\tilde{\Delta}^\beta)^2\bar{\sigma}
    \\
    \left.
    +\bar{h}\frac{d-2}{d^2}\tilde{\Delta}\bar{h}
    -\tilde{h}\tilde{\Delta}^{\alpha\beta}\tilde{h}
    \right],
\end{align}
where
\begin{equation}
    \tilde{\Delta}^\beta=\tilde{\Delta}-\beta\Delta.
\end{equation}
The resulting Gaussian integral is
\begin{align}
    \nonumber
    &J\int Dh^{TT}_{\mu\nu}D\xi_\mu D\bar{\sigma}D\bar{h}D\tilde{h}\,
    e^{-\delta^2_{\text{OnShell}}(S+S_{\text{gf}})}
    \\
    \nonumber
    =&\,
    i^{\frac d2}
    \left(
        \prod_{n=2}^{\infty}\prod_{m=1}^{D_n(d,2)}
        \frac{1}{\sqrt{\lambda''_n}}
    \right)
    \left(
        \prod_{n=2}^{\infty}\prod_{m=1}^{D_n(d,1)}
        \sqrt{\frac{\alpha}{\lambda'_n}}
    \right)
    \left(
        \prod_{n=2}^{\infty}\prod_{m=1}^{D_n(d,0)}
        \sqrt{\frac{\alpha\beta d^2}{2(d-2)(\lambda_n^\beta)^2}}
    \right)
    \\
    &\times
    \left(
        \prod_{n=0}^{1}\prod_{m=1}^{D_n(d,0)}
        \sqrt{\frac{d-2}{2d\lambda_n^{\alpha\beta}}}
    \right),
\end{align}
with
\begin{equation}
    \lambda''_n=\lambda_n^2+\frac{2R}{d(d-1)},
    \qquad
    \lambda_n^\beta=\tilde{\lambda}_n-\beta\lambda_n^0,
    \qquad
    \lambda_n^{\alpha\beta}
    =
    \frac{1}{d^2}
    \left(
        2\frac{\beta^2}{\alpha}\lambda_n^0-(d-2)\tilde{\lambda}_n
    \right).
\end{equation}

The ghost field is decomposed into transverse and longitudinal pieces,
\begin{equation}
    c^\mu=c^{T\mu}+\nabla^\mu c,
\end{equation}
with measure
\begin{equation}
    D\bar c^\mu Dc^\mu
    =
    \left(
        \prod_{n=1}^{\infty}\prod_{m=1}^{D_n(d,0)}
        \frac{1}{\lambda_n^0}
    \right)
    D\bar c^{T\mu}Dc^{T\mu}D\bar c Dc\,.
\end{equation}
The ghost action is
\begin{equation}
    S_{\text{gh}}
    =
    \int d^dx \sqrt{g}
    \left[
        \bar c^T_\mu\left(\frac{\Delta_1}{\sqrt\alpha}\right)c^{T\mu}
        +\bar c\left(\frac{2}{\sqrt\alpha d}\Delta\tilde\Delta^\beta\right)c
    \right].
\end{equation}
Hence
\begin{equation}
    \int D\bar c^\mu Dc^\mu e^{-S_{\text{gh}}}=
    \left(
        \prod_{n=2}^{\infty}\prod_{m=1}^{D_n(d,1)}
        \frac{\lambda'_n}{\sqrt\alpha}
    \right)
    \left(
        \prod_{n=1}^{\infty}\prod_{m=1}^{D_n(d,0)}
        \frac{2\lambda_n^\beta}{\sqrt\alpha d}
    \right).
\end{equation}
Collecting all contributions one obtains the complete on-shell one-loop result,
\begin{align}\label{eq:onelooponshellnew}
    \nonumber
    &\int Dh_{\mu\nu}D\bar c_\mu Dc_\mu\,
    e^{-\delta^2_{\text{OnShell}}(S+S_{\text{gf}})-S_{\text{gh}}}
    \\
    &=
    i^{d+2}
    \left(
        \prod_{n=2}^{\infty}\prod_{m=1}^{D_n(d,2)}
        \frac{1}{\sqrt{\lambda''_n}}
    \right)
    \left(
        \prod_{n=2}^{\infty}\prod_{m=1}^{D_n(d,1)}
        \sqrt{\lambda'_n}
    \right)
    \left(\sqrt{\frac{d}{2R}}\right)
    \left((d+1)\sqrt{\frac{(d-2)R}{d(d-1)}}\right).
\end{align}
This expression is manifestly independent of the gauge parameters.

\subsection{Off-shell effective action and residual gauge dependence}

Away from the equations of motion, the quadratic action retains explicit
$\Omega$-dependent terms. After the second scalar decomposition introduced above,
the off-shell quadratic form becomes
\begin{align}
    \nonumber
    \delta^2\left(S+S_{\text{gf}}\right)
    &=
    \frac{1}{32\pi G_N}
    \int d^dx \sqrt{g}
    \Bigg[
    h^{TT}_{\mu\nu}\left(\Delta_2+2\Omega\right)h^{TT\,\mu\nu}
    +\xi_\mu\left(\frac{2}{\alpha}(\Delta_1)^2+4\Delta_1\Omega\right)\xi^\mu
    \\
    \nonumber
    &\quad
    +\bar{\sigma}
    \left(
        \frac{2}{\alpha d^2}\Delta(\tilde{\Delta}^\beta)^2
        +\Delta\tilde{\tilde{\Delta}}\Omega
    \right)\bar{\sigma}
    +2\left(1-2\frac{1+\beta}{d}\right)
    \bar{\sigma}
    \left(
        \frac{\tilde{\Delta}\Delta}{\tilde{\Delta}^\beta}\Omega
    \right)i\bar h
    \\
    &\quad
    +\bar h
    \left(
        \frac{d-2}{d^2}\tilde{\Delta}
        -\frac{\tilde{\Delta}\tilde{\tilde{\Delta}}^\beta}{d(\tilde{\Delta}^\beta)^2}\Omega
    \right)\bar h
    +\tilde h
    \left(
        -\tilde{\Delta}^{\alpha\beta}+\frac{d-2}{d}\Omega
    \right)\tilde h
    \Bigg],
\end{align}
where
\begin{equation}
    \tilde{\tilde{\Delta}}=\Delta-\frac{2R}{d},
    \qquad
    \tilde{\tilde{\Delta}}^\beta
    =
    2\beta^2\Delta-(d-2)\tilde{\Delta}.
\end{equation}
With the shift
\begin{equation}
    \bar{\bar{\sigma}}
    =
    \bar{\sigma}
    -\frac{
        \left(1-2\frac{1+\beta}{d}\right)\tilde{\Delta}\Delta\,\Omega
    }{
        \tilde{\Delta}^\beta
        \left(
            \frac{2}{\alpha d^2}\Delta(\tilde{\Delta}^\beta)^2
            +\Delta\tilde{\tilde{\Delta}}\Omega
        \right)
    }i\bar h,
\end{equation}
and the rescalings
\begin{align}
    h^{TT}_{\mu\nu}&\to \sqrt{32\pi G_N}\,h^{TT}_{\mu\nu},
    &
    \xi_\mu&\to \sqrt{\frac{16\pi G_N}{\Delta_1}}\,\xi_\mu,
    \\
    \bar{\bar{\sigma}}&\to \sqrt{\frac{32\pi G_N d}{\Delta\tilde{\Delta}}}\,\bar{\bar{\sigma}},
    &
    \bar h,\tilde h&\to \sqrt{\frac{64\pi G_N d}{d-2}}\,\bar h,\tilde h,
\end{align}
the quadratic action is written as
\begin{equation}\label{eq:quadraticpartnew}
    \delta^2\left(\Gamma_k+S_{\text{gf}}\right)
    =
    \int d^dx \sqrt{g}\,
    \left[
        h^{TT}_{\mu\nu}\Gamma_{TT}h^{TT\,\mu\nu}
        +\xi_\mu\Gamma_{\xi\xi}\xi^\mu
        +\bar{\bar{\sigma}}\Gamma_{\sigma\sigma}\bar{\bar{\sigma}}
        +\tilde h\Gamma_{\tilde h\tilde h}\tilde h
        +\bar h\Gamma_{\bar h\bar h}\bar h
    \right],
\end{equation}
with
\begin{align}
    \Gamma_{TT}&=\Delta_2+2\Omega,\\
    \Gamma_{\xi\xi}&=\frac{1}{\alpha}\Delta_1+2\Omega,\\
    \Gamma_{\sigma\sigma}&=\frac{2(\tilde{\Delta}^\beta)^2+\alpha d^2 \tilde{\tilde{\Delta}}\Omega}{\alpha d\tilde{\Delta}},\\
    \Gamma_{\tilde h\tilde h}&=\frac{2d}{2-d}\tilde{\Delta}^{\alpha\beta}+2\Omega,\\
    \Gamma_{\bar h\bar h}
    &=
    \frac{2}{d}\tilde{\Delta}
    -\frac{2\tilde{\Delta}\tilde{\tilde{\Delta}}^\beta\Omega}{(d-2)(\tilde{\Delta}^\beta)^2}
    +\frac{
        2d\left(1-2\frac{1+\beta}{d}\right)^2
        \tilde{\Delta}^2\Delta^2\Omega^2
    }{
        (d-2)(\tilde{\Delta}^\beta)^2
        \left(
            \frac{2}{\alpha d^2}\Delta(\tilde{\Delta}^\beta)^2
            +\Delta\tilde{\tilde{\Delta}}\Omega
        \right)
    }.
\end{align}

The ghost action can be factorised as $\det M=\sqrt{\det M'\det M''}$, with
\begin{equation}\label{eq:ghost1new}
    S'_{gh}
    =
    \int d^dx \sqrt{g}
    \left[
        \bar c^{\prime T}_\mu \mathcal Q'_{TT} c^{\prime T\,\mu}
        +\bar{\hat c}'\hat{\mathcal Q}'\hat c'
        +\bar{\tilde c}'\tilde{\mathcal Q}'\tilde c'
        +b^{\prime T}_\mu \mathcal Q'_{TT} b^{\prime T\,\mu}
        +\hat b' \hat{\mathcal Q}' \hat b'
        +\tilde b' \tilde{\mathcal Q}' \tilde b'
    \right],
\end{equation}
where
\begin{equation}
    \mathcal Q'_{TT}=\frac{\Delta_1}{\alpha},
    \qquad
    \hat{\mathcal Q}'=\frac{2}{\alpha d}\frac{(\tilde{\Delta}^\beta)^2}{\tilde{\Delta}},
    \qquad
    \tilde{\mathcal Q}'=\frac{2d}{2-d}\tilde{\Delta}^{\alpha\beta},
\end{equation}
and
\begin{equation}\label{eq:ghost2new}
    S''_{gh}
    =
    \int d^dx \sqrt{g}
    \left[
        \bar c^{\prime\prime T}_\mu \mathcal Q''_{TT} c^{\prime\prime T\,\mu}
        +\bar{\hat c}''\hat{\mathcal Q}''\hat c''
        +\bar{\tilde c}''\tilde{\mathcal Q}''\tilde c''
        +b^{\prime\prime T}_\mu \mathcal Q''_{TT} b^{\prime\prime T\,\mu}
        +\hat b''\hat{\mathcal Q}''\hat b''
        +\tilde b''\tilde{\mathcal Q}''\tilde b''
    \right],
\end{equation}
with
\begin{equation}
    \mathcal Q''_{TT}=\Delta_1,
    \qquad
    \hat{\mathcal Q}''=\frac{2}{d}\tilde{\Delta},
    \qquad
    \tilde{\mathcal Q}''=
    \frac{2}{\alpha}\frac{2-d}{d^3}\frac{(\tilde{\Delta}^\beta)^2}{\tilde{\Delta}^{\alpha\beta}}.
\end{equation}
The one-loop effective action is therefore gauge dependent off shell. On shell $S'_{gh}$ yields precisely the counterterms needed to cancel the
gauge-dependent part of the gauge-fixed action.

\bigskip

This section helps to identify the off-shell origin of the $\alpha$-dependence: it resides in the non-physical scalar, vector, and ghost operators. The one-loop calculation gives an explicit determinant-level realisation of the essential projection used in the flow analysis, which we perform in the next section.

\section{All-orders essential scheme}
\label{sec:allorders}

The previous sections suggest that the gauge dependence of the proper-time flow
can be confined to redundant off-shell structures. We now show how this logic
extends beyond a finite curvature truncation by promoting the field-redefinition
kernel to a curvature-dependent function. In this way, one can implement the
essential scheme to all orders in the scalar curvature~\cite{Codello:2007bd,Machado:2007ea,Falls:2016msz,Morris:2022btf,Falls:2024noj,Baldazzi:2023pep}.

We consider the ansatz of the field dependent field redefinition
\begin{equation}
    \Psi_{\mu\nu}=\gamma(R)\,g_{\mu\nu}.
\end{equation}
The generalized proper-time flow then becomes
\begin{dmath}\label{eq:allordersmaster}
\int d^dx \sqrt{g}
\left[
\frac{k^d}{8\pi}\partial_t \tilde\rho
+\frac{k^{d-2}}{16\pi \tilde G^2}R\,\partial_t \tilde G
-\frac{d\,k^{d-2}}{16\pi\tilde G}(2+\gamma(R))\Omega
+k^{d-4}\Xi\left((d-4)\tilde\vartheta+\partial_t\tilde \vartheta\right)
\right]
=
\sum_{n=2}^{\infty} D_n(d,2)W\!\bigl(\Gamma_{TT}(n,2)\bigr)
+\sum_{n=2}^{\infty} D_n(d,1)W\!\bigl(\Gamma_{\xi\xi}(n,1)\bigr)
+\sum_{n=2}^{\infty} D_n(d,0)W\!\bigl(\Gamma_{\sigma\sigma}(n,0)\bigr)
+\sum_{n=2}^{\infty} D_n(d,0)W\!\bigl(\Gamma_{\bar h\bar h}(n,0)\bigr)
+\sum_{n=0}^{1} D_n(d,0)W\!\bigl(\Gamma_{\tilde h\tilde h}(n,0)\bigr)
-\sum_{n=1}^{\infty} D_n(d,1)W\!\bigl(\mathcal Q'_{TT}(n,1)\bigr)
-\sum_{n=1}^{\infty} D_n(d,1)W\!\bigl(\mathcal Q''_{TT}(n,1)\bigr)
-\sum_{n=2}^{\infty} D_n(d,0)W\!\bigl(\hat{\mathcal Q}'(n,0)\bigr)
-\sum_{n=2}^{\infty} D_n(d,0)W\!\bigl(\hat{\mathcal Q}''(n,0)\bigr)
-D_1(d,0)W\!\bigl(\tilde{\mathcal Q}'(1,0)\bigr)
-D_1(d,0)W\!\bigl(\tilde{\mathcal Q}''(1,0)\bigr).
\end{dmath}
Here the components $\Gamma(n,s)$ and $\mathcal Q(n,s)$ are the eigenvalues
of the operators appearing in \eqref{eq:quadraticpartnew},
\eqref{eq:ghost1new}, and \eqref{eq:ghost2new}, expressed in terms of the
eigenvalues $\lambda_n^s$. 
The topological term is incorporated using its
flow in the more general background, eq.~\eqref{eq:betatheta_sec3}, at the
Gaussian fixed point. The fixed value of $\tilde{\rho}$ is evaluated exactly by taking the trace to the lowest order in the heat kernel expansion. For the remaining part of the flow, we approximate the spectral sum up to a sufficiently large n, ensuring that the contribution of additional terms is negligible.\\
It is also interesting to study the flow to a finite order of $R$. Here, imposing the constraint $\beta=\alpha (d-2)/2$ diagonalises the Hessian \eqref{eq:hessianTTnew}, allowing the trace to be computed directly via the heat kernel expansion with ease. In contrast, the constraint is unnecessary when using the spectral sum, so, in the $R^\infty$ truncation, the flow contains both $\alpha$ and $\beta$, and we can achieve independence from both parameters.

The essential scheme is implemented by fixing the inessential coupling,
\begin{equation}
    \partial_t\tilde\rho=0.
\end{equation}
One then solves simultaneously for $\partial_t\tilde G$ and $\gamma(R)$.
The key structural point is that these quantities play different roles:
$\partial_t\tilde G$ contains only on-shell information and is independent of
the gauge parameters at all curvature orders, while $\gamma(R)$ is an
off-shell quantity carrying the full gauge dependence.\\
If we consider the flow truncated up to $R^N$, $\partial_t \tilde{G}$ is gauge independent up to $\tilde{G}^{N+1}$. At order $N=5$ and $d=4$
\begin{dmath}
    \partial_t\tilde G=2\tilde G-\frac{8}{3\pi }\frac{\tilde G^2}{m-1}-\frac{29}{40\pi^2}\frac{\tilde G^3}{(m-2)(m-1)}-\frac{2459}{68040 \pi^3}\frac{m \tilde G^4}{ (m-2)^2 (m-1)^2}+\frac{5441}{3265920 \pi ^4}\frac{m (m+1)\tilde G^5}{(m-2)^3 (m-1)^3}+\frac{39059}{53887680 \pi ^5}\frac{m(m+1)(m+2)\tilde{G}^6}{(m-2)^4 (m-1)^4}+O(\tilde{G}^7),
\end{dmath}
and $\gamma(R)$ is gauge dependent. For $m=3$:
\begin{dmath}
    \gamma(0)=\frac{1}{4} \left(-\frac{(\alpha -3)^2}{\alpha ^2}-(\alpha -3)^2+\frac{4 \pi ^2 (\alpha -3)^4}{\alpha ^2 (2 \pi  (\alpha -3)+\alpha  \tilde G)^2}+\frac{4 \pi ^2 (\alpha -3)^4}{(2 \pi  (\alpha -3)+\tilde G)^2}-\frac{48 \pi }{\alpha  \tilde G-4 \pi }-\frac{80 \pi }{\tilde G-4 \pi }-32\right).
\end{dmath}

After the shift\footnote{This shift is equivalent to $k\to m k$ in the profile function \eqref{eq:PT_profile}. In the limit $m\to\infty$, it takes the exponential form $W(z)=\exp(-z/k^2)$.} $\tilde G\to m \tilde{G}$, in the limit $m\to\infty$, the beta function
\begin{equation}
    \partial_t\tilde G=2 \tilde G-\frac{8 \tilde G^2}{3 \pi }-\frac{29 \tilde G^3}{40 \pi ^2}-\frac{2459 \tilde G^4}{68040 \pi ^3}+\frac{5441 \tilde G^5}{3265920 \pi ^4}+\frac{39059 \tilde G^6}{53887680 \pi ^5}+O(\tilde{G}^7)
\end{equation}
consistently reproduces the same result as~\cite{Falls:2024noj}. At this order $O(\tilde{G}^7)$ is gauge dependent, and the full gauge independence is restored at all curvature orders.

\bigskip

We now turn our attention to the fixed point structure.

\begin{figure}[h]
    \centering
 \includegraphics[width=.5\linewidth]{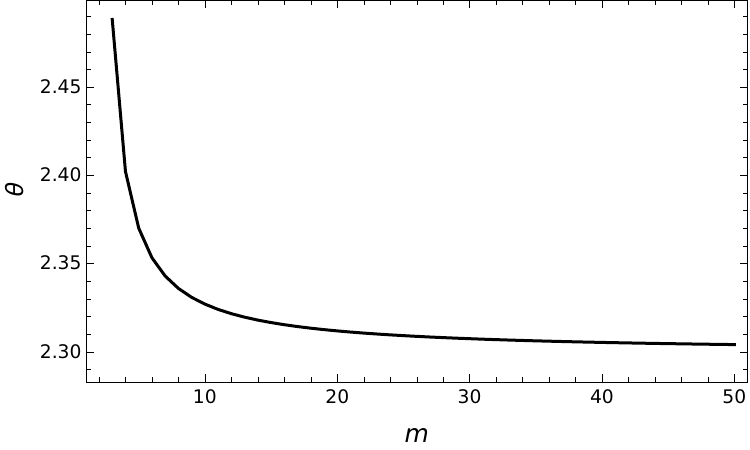}
    \caption{The critical exponent $\theta$ for the spherical background in $d=4$ dimensions at the non-Gaussian fixed point for different values of $m$.}
    \label{critical exponent sphere plot true}
\end{figure}

In $d=4$, the all-orders flow of $\tilde G$ exhibits a non-Gaussian fixed
point,
\begin{equation}
    \beta_{\tilde G}(\tilde G_*)=0,
\end{equation}
with critical exponent
\begin{equation}
    \theta
    =
    -\left.
    \frac{\partial \beta_{\tilde G}}{\partial \tilde G}
    \right|_{\tilde G=\tilde G_*}.
\end{equation}
As shown in Table \ref{tab:critical_exponents}, within the present approximation, the critical exponent in the $R^N$ truncation converges to the corresponding all-orders value, which depends only mildly on the proper-time regulator parameter $m$.
This dependence is plotted in fig.~\ref{critical exponent sphere plot true}.
When the truncation is restricted to $R^3$, the beta function and the critical exponents match the results from eq. \eqref{eq:betaGess_sec3} with the same truncation. This provides strong evidence that the residual gauge dependence is progressively removed upon the inclusion of higher curvature orders, independently of the background symmetries.
\begin{table}[h]
\centering
\begin{tabular}{|c|c|c|c|c|c|}
\hline
$O(R^N)$ & $\theta(m=3)$ & $\theta(m=4)$ & $\theta(m=8)$ & $\theta(m=20)$ & $\theta(m=100)$\\
\hline
$R$  & $2$ & $2$ & $2$    & $2$  & $2$\\
$R^2$ & $2.475$ & $2.394$ & $2.331$ & $2.308$ & $2.298$\\
$R^3$  & $2.537$ & $2.433$ & $2.354$ & $2.327$ & $2.315$\\
$R^4$  & $2.527$ & $2.429$ & $2.353$    & $2.326$ & $2.314$\\
$R^5$  & $2.509$ & $2.424$ & $2.352$  & $2.325$  & $2.314$\\
$R^\infty$  & $2.488$ & $2.422$ & $2.352$  & $2.325$  & $2.314$\\
\hline
\end{tabular}
\caption{Critical exponents at different orders of the curvature expansion for different values of $m$.}
    \label{tab:critical_exponents}
    \end{table}
    
In gauges where the kernel is regular,
one may also evaluate
\begin{equation}
    \gamma_*(\tilde R)
    =
    \gamma(k^2\tilde R)\big|_{\tilde G=\tilde G_*}.
\end{equation}
This quantity retains the full gauge dependence of the off-shell completion,
even though the fixed-point condition for $\tilde G$ does not.

The curvature-dependent kernel also governs the induced renormalisation group
flow of the metric itself,
\begin{equation}
    \partial_t g_{\mu\nu}=\Psi_{\mu\nu}.
\end{equation}
Setting
\begin{equation}
    g_{\mu\nu}=\omega_k \hat g_{\mu\nu},
\end{equation}
with $\omega_k=R^{-1}$ and $\hat g_{\mu\nu}$ a metric of unit curvature, one
finds at the non-Gaussian fixed point
\begin{equation}\label{eq:omega_flow}
    \partial_t \tilde\omega_k
    =
    \tilde\omega_k
    \left(
        2+\gamma_*(\tilde\omega_k^{-1})
    \right),
\end{equation}
where
\begin{equation}
    \tilde\omega_k=k^2\omega_k.
\end{equation}
This equation makes clear that the gauge-dependent kernel governs the off-shell
RG parametrisation of the background geometry, while the gauge-independent flow
of Newton's coupling belongs to the essential sector.

\bigskip

The gauge-fixed proper-time flow is not gauge
independent off shell.  Once the field-redefinition kernel is promoted
to a curvature-dependent function, the full gauge dependence can be shifted into
$\gamma(R)$, leaving a gauge-independent flow for the essential Newton
coupling.

\begin{figure}[h]
    \centering    \includegraphics[width=.5\linewidth]{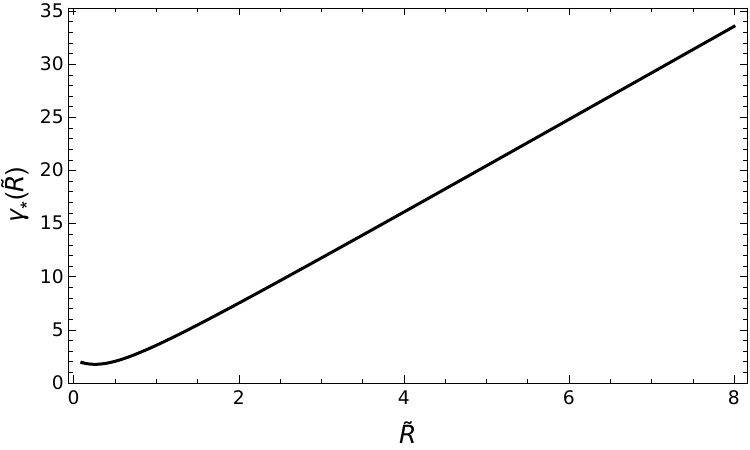}
    \caption{$\gamma_*(\tilde{R)}$ in $d=4$ dimensions for $\alpha=0$, $\beta=1$, and $m=8$.}
    \label{gamma function alpha=0 beta=1 true}
\end{figure}

\begin{figure}[h]
    \centering
    \includegraphics[width=.5\linewidth]{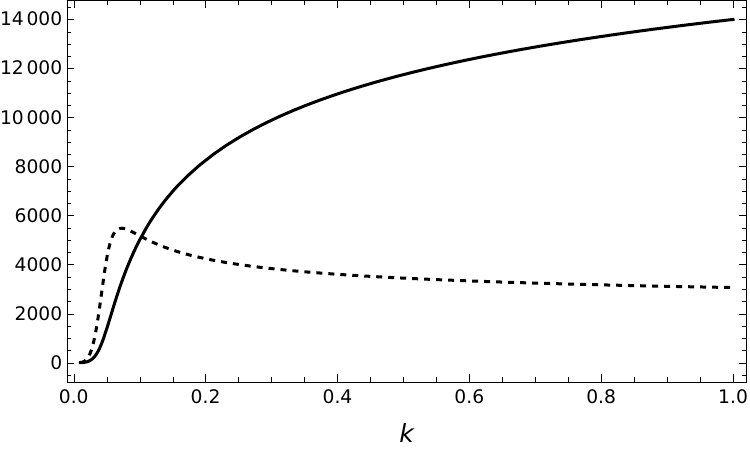}
    \caption{$\tilde{\omega}_k$ (solid line) and $\partial_t \tilde{\omega}_k$ (dashed line) in $d=4$ dimensions for $\alpha=0$, $\beta=1$, and $m=8$, with the boundary $\tilde{\omega}_{1}=1.4\times10^{4}$, where we take units such that some reference scale $k_0 =1$.}
    \label{omega function alpha=0 beta=1 true}
\end{figure}

This identifies a clean division between the physical and unphysical parts of
the flow. The beta function of $\tilde G$, together with its non-Gaussian
fixed point and critical exponent, belongs to the essential on-shell sector.
By contrast, $\gamma(R)$ parametrises the redundant off-shell completion and
therefore carries the gauge dependence. In this way, the all-orders essential
scheme provides the natural continuation of both the truncated proper-time
analysis and the explicit on-shell cancellation mechanism discussed above. An example of $\gamma_\star(\tilde{R})$ as a function of $\tilde{R} = R/k^2$ is plotted in fig.~\ref{gamma function alpha=0 beta=1 true} in the gauge with $\alpha = 0$ and $\beta =1$. Additionally, we can integrate \eqref{eq:omega_flow} to find the field redefinition as a function of RG scale at the fixed point. The form of $\tilde{\omega}_k$ and its derivative are given in fig.~\ref{omega function alpha=0 beta=1 true} for $\alpha = 0$ and $\beta =1$.


\section{Conclusions}
\label{sec:conclusions}

In this work, we have analysed gauge dependence in proper-time renormalisation group flows for asymptotically safe quantum gravity from the viewpoint of essential couplings. The central result is that the gauge dependence of the proper-time flow is an off-shell effect: it is tied to redundant directions in theory space and can be removed from the physical running by organising the flow in an essential scheme. Thus, the relevant criterion is not whether every off-shell intermediate quantity is gauge independent, but whether the projected essential sector is. Ultimately, the essential scheme is not essential to obtain physical results. However, when approximations are made, it helps to clean the truncated equations of nonphysical information, which can cross-contaminate the physical critical exponents of gauge-invariant operators. It is also worth stressing that an appropriate gauge-dependent regularisation, particularly in the ghost sector, is required in order to cancel the gauge dependence of the regularised theory. Thus, to some extent, the gauge independence is achieved by hand using the freedom to pick the regularisation appropriately. 

We implemented this idea by supplementing the proper-time flow with scale-dependent field redefinitions. Retaining terms with up to four derivatives in the flow equation, the field-redefinition kernel $\Psi_{\mu\nu}$ absorbs contributions proportional to the equations of motion and fixes the running along inessential directions. In particular, we used this freedom to impose a renormalisation condition on the inessential cosmological coupling. The resulting flow separates the off-shell completion, which may depend on the gauge fixing, from the running of the essential couplings.

Within this scheme, the full beta function of the dimensionless Newton coupling still depends explicitly on the gauge parameter to all orders in $G$. After projection onto the essential on-shell sector this dependence cancels up to the order considered when also truncating the beta function to cubic order in $G$. The gauge-independent beta function exhibits a non-Gaussian ultraviolet fixed point in four dimensions for a broad range of proper-time regulators. The associated critical exponent remains positive, while the residual regulator dependence is mild. This shows that the fixed-point structure found in the proper-time flow is not simply an artifact of a gauge-dependent off-shell truncation, but survives in the sector expected to carry the physical information.

We also checked the determinant-level origin of this mechanism in a fixed one-loop calculation. On maximally symmetric backgrounds, the gauge-fixed fluctuation operator and the ghost determinants are separately gauge dependent away from the equations of motion. On shell the gauge-dependent contributions from the non-physical fluctuation sectors are cancelled by the ghost sector, leaving a gauge-independent determinant. This explicit cancellation gives a determinant-level realisation of the essential projection used in the flow analysis and clarifies the role of the ghost split in the proper-time regularisation.

The construction extends naturally beyond a finite curvature truncation. By promoting the field-redefinition kernel to a curvature-dependent function $\gamma(R)$, the off-shell gauge dependence can be shifted into the redundant kernel, while the flow of Newton's coupling is determined by on-shell data. At finite order in the curvature expansion, the beta function is gauge independent up to the corresponding order in $\tilde G$; in the all-orders limit, this reproduces the essential flow found in the on-shell perturbative approach. In this sense, the proper-time formalism can isolate universal information, provided the flow is formulated in terms of essential rather than off-shell couplings.

These results support the broader view that asymptotic safety should be formulated, as far as possible, in terms of essential and on-shell data~\cite{Baldazzi:2021ydj,Baldazzi:2021orb,Knorr:2022ilz,Knorr:2023usb,Falls:2024noj,Benedetti:2011ct,Falls:2017cze}. They also suggest that part of the apparent gauge, parametrisation, and scheme dependence encountered in functional approaches to quantum gravity reflects the use of redundant off-shell representatives rather than an ambiguity in the physical content of the theory~\cite{Gies:2015tca,Nink:2014yya,Ohta:2016npm,Ohta:2016jvw,Bonanno:2020bil,Donoghue:2019clr}.

Several approximations remain. The analysis was carried out within a proper-time framework and in restricted truncations, so it does not establish gauge independence of the full non-perturbative theory. Larger truncations, higher-derivative operators, more general backgrounds, and different regulator choices may introduce additional subtleties. It will therefore be important to test whether the same essential projection continues to isolate a gauge-independent sector in comparisons with lattice approaches such as causal dynamical triangulations~\cite{Hamber:2009mt,Laiho:2016nlp,Reuter:2011ah,Loll:2019rdj,Ambjorn:2012jv,Ambjorn:2020rcn,Ambjorn:2024qoe,Schiffer:2025cqc}.

A particularly useful next step would be a unified treatment of gauge, parametrisation, and regulator dependence within the same essential framework. Since all three are tied to the distinction between physical and redundant structures, such an analysis could sharpen the interpretation of fixed points and critical exponents in asymptotically safe gravity and provide more criteria for identifying universal quantities.

In this work, we have used a one-loop improved scheme. In this setting, BRST symmetry is controlled by the standard gauge fixing ansatz without the need to renormalise the ghost action or the gauge fixing. Going beyond the current approximation will involve correctly incorporating the machinery of BRST cohomology to track the running of the gauge parameters.

Finally, another direction is to combine the present essential scheme with the flow of relational (i.e., gauge invariant) observables developed in \cite{Baldazzi:2021fye, Thiemann:2024vjx, Ferrero:2024rvi, Ferrero:2025efd}.  Extending and adapting the present mechanism to the flow of relational observables would clarify whether the scaling dimensions of relational observables are insensitive to gauge choices once the essential projection is imposed.

\section*{Acknowledgments}
We would like to thank Alfio Bonanno and Oleg Melichev for inspiring discussions. K.F. acknowledges financial support from ANII-SNI-2022-1-1012554.

\nocite{apsrev42Control}
\bibliographystyle{apsrev4-2}
\bibliography{bib}
\end{document}